\documentclass{article}

\usepackage{setspace}
\usepackage{color}
\usepackage{graphicx}
\usepackage[margin=1in]{geometry}
\usepackage{breakcites}
\usepackage{titling}
\usepackage[bottom]{footmisc}
\pdfoutput=1

\title{Brain State Flexibility Accompanies Motor-Skill Acquisition}
\author{Pranav G. Reddy $^1$, Marcelo G. Mattar $^2$, Andrew C. Murphy $^{1,3}$, Nicholas F. Wymbs $^4$\\ Scott T. Grafton $^5$, Theodore D. Satterthwaite $^6$, Danielle S. Bassett $^{1,7,8}$}
\date{25 January 2017}

\begin{document}
\raggedbottom
\footnotetext[1]{Department of Bioengineering, University of Pennsylvania, Philadelphia, PA 19104 USA}
\footnotetext[2]{Department of Psychology, University of Pennsylvania, Philadelphia, PA 19104, USA}
\footnotetext[3]{Perelman School of Medicine, University of Pennsylvania, Philadelphia, PA 19104, USA}
\footnotetext[4]{Department of Physical Medicine and Rehabilitation, Johns Hopkins University, Baltimore, MD 21218 USA}
\footnotetext[5]{Department of Psychological and Brain Sciences, University of California, Santa Barbara, CA 93106 USA}
\footnotetext[6]{Department of Psychiatry, University of Pennsylvania, Philadelphia, PA 19104, USA}
\footnotetext[7]{Department of Electrical and Systems Engineering, University of Pennsylvania, Philadelphia, PA 19104 USA}
\footnotetext[8]{To whom correspondence should be addressed: dsb@seas.upenn.edu.}

\maketitle
\pagebreak

\begin{abstract}
Learning requires the traversal of inherently distinct cognitive states to produce behavioral adaptation. Yet, tools to explicitly measure these states with non-invasive imaging -- and to assess their dynamics during learning -- remain limited. Here, we describe an approach based on a novel application of graph theory in which points in time are represented by network nodes, and similarities in brain states between two different time points are represented as network edges. We use a graph-based clustering technique to identify clusters of time points representing canonical brain states, and to assess the manner in which the brain moves from one state to another as learning progresses. We observe the presence of two primary states characterized by either high activation in sensorimotor cortex or high activation in a frontal-subcortical system. Flexible switching among these primary states and other less common states becomes more frequent as learning progresses, and is inversely correlated with individual differences in learning rate. These results are consistent with the notion that the development of automaticity is associated with a greater freedom to use cognitive resources for other processes. Taken together, our work offers new insights into the constrained, low dimensional nature of brain dynamics characteristic of early learning, which give way to less constrained, high-dimensional dynamics in later learning.
\end{abstract}

\section*{Keywords:} motor sequence learning, graph theory, discrete sequence production, brain state flexibility


\section*{Introduction}

The human brain is an inherently adaptive \cite{mattar2016varying}, plastic \cite{della2015sensorimotor} organ. Its fundamental malleability supports changes to its architecture and function that are advantageous to human survival. Importantly, such changes can occur on multiple time scales: from the long time scales of evolution \cite{kirschner1998evolvability,clune2013evolutionary} to the shorter time scales of multi-year development \cite{gu2015emergence}, or even short-term learning \cite{ellefsen2015neural,hermundstad2011learning}. Notably, even in the shortest time scales of learning, adaptation can occur over multiple spatial scales \cite{mattar2016brain}, from the level of single neurons \cite{richardson2012activity} to the level of large-scale systems \cite{tunik2007beyond}. Moreover, this adaptation can affect functional dynamics \cite{heitger2012motor,krakauer2005adaptation,krakauer2004differential,grefkes2004human} or can evoke a direct change in the structure of neuroanatomy, driving new dendritic spines \cite{xu2009rapid}, axon collaterals \cite{chkovskii2004cortical}, and myelination \cite{sampaio2013motor}. 

Malleability, adaptability, and plasticity often manifest as a variability in quantitative statistics that describe the structure or function of a system. In the large-scale human brain, such statistics can include measures of neurophysiological noise \cite{garrett2014brain,garrett2013moment,breakspear2011networks} or changes in patterns of resting state functional connectivity \cite{deco2009key,deco2011emerging,deco2013resting}. More recently, dynamic reconfiguration of putative functional modules in the brain -- groups of functionally connected areas identified using community detection algorithms \cite{porter2009,fortunato2010} -- has been used to define a notion of network flexibility \cite{bassett2011dynamic}, which differs across individuals and is correlated with individual differences in learning \cite{bassett2011dynamic}, cognitive flexibility \cite{braun2015dynamic}, and executive function \cite{braun2015dynamic}. 

\begin{figure}
\begin{center}
\includegraphics[width=95mm]{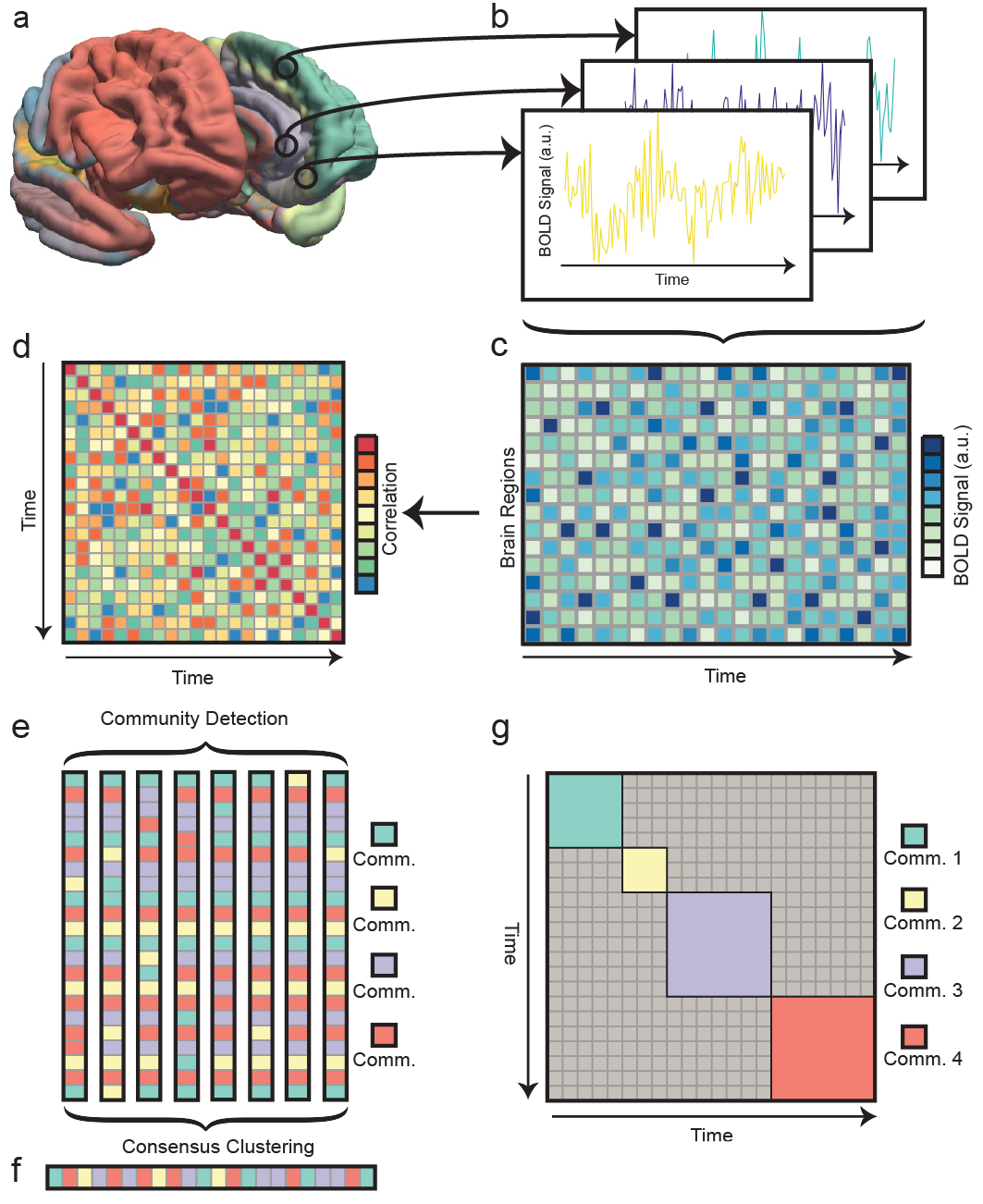}
\end{center}
\caption{\textbf{Schematic Depicting Construction of Adjacency Matrices.} \emph{(a)} Blood-oxygen-level-dependent (BOLD) signal from functional magnetic resonance imaging (fMRI) data was acquired from healthy adult subjects. \emph{(b)} We calculated the mean BOLD magnitudes in each of 112 cortical and subcortical regions as a function of time. \emph{(c)} The regional time series is represented in matrix format, and \emph{(d)} the correlation between matrix columns (TRs) is used to create a time-by-time adjacency matrix. The $ij^{th}$ element of this matrix measures the similarity between the regional pattern of BOLD magnitude between TR $i$ and TR $j$. Adjacency matrices representing time-by-time networks form the fundamental data structure on which community detection algorithms function. We maximize a modularity quality function informed by these matrices to extract network communities: groups of TRs that show similar regional patterns of BOLD magnitudes. \emph{(e)} Due to the near-degeneracy of the modularity landscape, this procedure is repeated 100 times per matrix. \emph{(f)} Across these 100 partitions of TRs (nodes) into groups (communities), we construct a representative or ``consensus'' partition \emph{(g)} that summarizes the significant structure in the original matrix.  \label{fig8}}
\end{figure}

Indeed, in the context of motor skill learning, dynamic network techniques have proven to be particularly advantageous for longitudinal designs, where data is collected from the same participants at multiple time points interspersed throughout the learning process \cite{bassett2015learning,wymbs2015human,bassett2013task}. Using a 6-week longitudinal design where participants trained motor sequences while undergoing functional magnetic resonance imaging, motor system activity was found to be associated with both increasing and decreasing motor system activity, with sequence-specific representations varying across multiple distinct timescales \cite{wymbs2015human}. With a network modeling approach based on coherent activity between brain regions, the same dataset revealed the existence of a core-periphery structure that changes over the course of training and predicts individual differences in learning success \cite{bassett2013task}. And more recently, these changes were shown to reflect a growing autonomy between sensory and motor cortices, and the release of cognitive control hubs in frontal and cingulate cortices \cite{bassett2015learning}. Yet despite these promising advances, dynamic network reconfiguration metrics are fundamentally unable to assess changes in the patterns of \emph{activity} that are characteristic of brain dynamics, as they require the computation of functional connectivity estimates over extended time windows \cite{telesford2016detection,bassett2013robust}.
 
To overcome this weakness, we developed an alternative technique inspired by network science to identify temporal activation patterns and to assess their flexibility \cite{medaglia2015flexible,chen2016large}. Leveraging the same longitudinal dataset from the above studies, we begin by defining a brain state as a pattern of regional activity -- for instance, estimated from functional magnetic resonance imaging (fMRI) -- at a single time point \cite{gu2015controllability,liu2013decomposition, liu2013time} (Fig.~\ref{fig8}). Time points with similar activity patterns are then algorithmically clustered using a graph-based clustering technique \cite{porter2009,fortunato2010}, producing sets of similar brain states. Finally, by focusing on the transitions from one state to another, we estimate the rate of switching between states. This approach is similar to techniques being concurrently developed in the graph signal processing literature \cite{huang2016graph,goldsberry2016brain}, and it allows us to ask how activation patterns in the brain change as a function of learning. We address this question in the context of the explicit acquisition of a novel motor-visual skill, which is a quintessential learning process studied in both human and animal models. As participants practice the task, we hypothesize that the brain traverses canonical states differently, that characteristics of this traversal predict individual differences in learning, and that the canonical states themselves are inherently different in early \emph{versus} late learning.

\begin{figure}
\centering
\includegraphics[width=90mm]{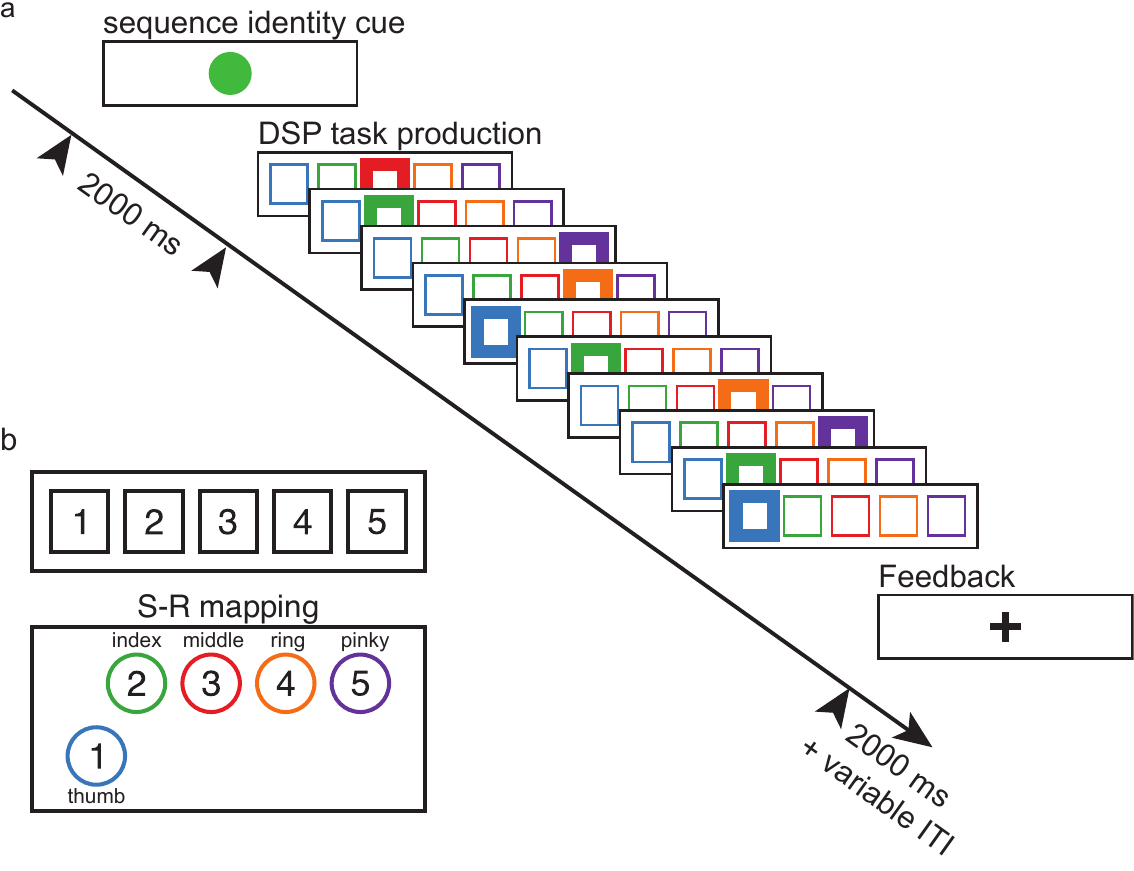}
\caption{\textbf{Trial Structure.} \emph{(a)} Each trial began with the presentation of a sequence-identity cue that remained on screen for 2 s. Each of the six trained sequences was paired with a unique identity cue. A discrete sequence-production (DSP) event structure was used to guide sequence production. The onset of the initial DSP stimulus (thick square, colored red in the task) served as the imperative to produce the sequence. A correct key press led to the immediate presentation of the next DSP stimulus (and so on) until the ten-element sequence was correctly executed. Participants received a "+" as feedback to signal that a sequence was completed and to wait (approximately 0-6 s) for the start of the next trial. This waiting period was called the "intertrial interval" (ITI). At any point, if an incorrect key was hit, a participant would receive an error signal (not shown in the figure), and the DSP sequence would pause until the correct response was received. \emph{(b)} There was direct S-R mapping between a conventional keyboard or an MRI-compatible button box (lower left) and a participant's right hand, so that the leftmost DSP stimulus cued the thumb and the rightmost stimulus cued the pinky finger. Note that the button location for the thumb was positioned to the lower left for maximum comfort and ease of motion.} \label{fig1}
\end{figure}

To test these hypotheses, 20 healthy adult human participants practiced a set of ten-element motor sequences. Execution of the sequences involved the conversion of a visual stimulus to a motor response in the form of a button press (see \textbf{Fig.~\ref{fig1}}). BOLD data was acquired during task performance on 4 separate occasions each two weeks apart (Fig.~\ref{fig8}a-b); between each consecutive pair of scanning sessions, participants practiced the sequences at home in approximately ten home training sessions. To assess behavioral change, we defined movement time (MT) to be the time between the first button press and the last button press for any given sequence, and learning rate was quantified by the exponential drop-off parameter of a two-term exponential function fit to the MT data. To assess the change in brain activity related to behavioral change, we divided the brain into 112 cortical and subcortical regions, and calculated regional BOLD time series (Fig.~\ref{fig8}c). We defined a brain state to be a pattern of BOLD magnitudes across regions, at each time point. We then quantified the similarities in brain states across time with a rank correlation measure between all pairs of states to create a symmetrical correlation matrix for each trial block (Fig.~\ref{fig8}d). Within each trial block we used network-based clustering algorithms to find recurrent brain states independent of their temporal order (Fig.~\ref{fig8}e-f). 

We observed three to five communities -- brain states, with two anti-correlated ``primary states'' occurring more frequently than the rest. We also observed that ``state flexibility'' -- the flexible switching among all brain states -- increases with task practice, being largely driven by contributions from brain regions traditionally associated with task learning and memory. Moreover, individuals with higher state flexibility learned faster than individuals with less switching between brain states. These results demonstrate that the global pattern of brain activity offers important insights into neurophysiological dynamics supporting adaptive behavior, underscoring the utility of a state-based assessment of whole brain dynamics in understanding higher order cognitive functions such as learning.

\section*{Materials and Methods}

\subsection*{Experiment and Data Collection.} \textit{Ethics statement}. In accordance with the guidelines set out by the Institutional Review Board of the University of California, Santa Barbara, twenty-two right-handed participants (13 females and 9 males) volunteered to participate and provided informed consent in writing. Separate analyses of the data acquired in this study are reported elsewhere \cite{bassett2015learning,wymbs2015human,bassett2013task}. 

\textit{Experimental Setup and Procedure}. Head motion was calculated for each subject as mean relative volume-to-volume displacement. Two participants were excluded from the following analyses: one failed to complete the entirety of the experiment and the other had persistent head motion greater than 5 mm during MRI scanning. The 20 remaining participants all had normal or corrected vision and none had any history of neurological disease or psychiatric disorders. In total, each participant completed at least 30 behavioral training sessions over the course of 6 weeks, a pre-training fMRI session and three test fMRI sessions. The training used a module that was installed on the participant's laptop by an experimenter. Participants were given instructions on how to use the module and were required to train at minimum ten days out of each of 3 fourteen day periods. Training began immediately after the pre-training fMRI session and test scans were conducted approximately fourteen days after each previous scan (during which training also took place). Thus a total of 4 scans were acquired over the approximately 6 weeks of training.

\textit{Training and trial procedure}. Participants practiced a set of ten-element sequences in a discrete sequence-production (DSP) task, which required participants to generate these responses to visual stimuli by pressing a button on a laptop keyboard with their right hand (see \textbf{Figure \ref{fig1}}).  Sequences were represented by a horizontal array of five square stimuli, where the thumb corresponded to the leftmost stimuli and the pinky corresponded to the rightmost stimuli. The imperative stimulus was highlighted in red and the next square to be pressed in the sequence was highlighted immediately after a correct key press. The sequence only continued once the appropriate key was pressed. Participants had an unlimited amount of time to complete each trial, and were encouraged to remain accurate rather than swift.

Each participant trained on the same set of six different ten element sequences, with three different levels of exposure: extensively trained (EXT) sequences that were practiced for 64 trials each, moderately trained (MOD) sequences that were practiced for 10 trials each, and minimally trained (MIN) sequences that were practiced for 1 trial each. Sequences included neither repetitions ("11", for example) nor patterns such as trills ("121", for example) or runs ("123", for example). All trials began with a sequence-identity cue, which informed participants which sequence they would have to type. Each identity cue was associated with only a single sequence and was composed of a unique shape and color combination. EXT sequences, for example, were indicated by a cyan or magenta circle, MOD sequences by a red or green triangle, and MIN sequences by orange or white stars. Participants reported no difficulty viewing the identity cues. After every set of ten trials, participants were given feedback about the number of error-free sequences produced and the mean time to produce an error-free sequence.

Each test session in a laboratory environment was completed after approximately ten home training sessions (over the course of fourteen days) and each participant took part in three test sessions, not including the pre-training session, which was identical to the training sessions. To familiarize the participants with the task, we introduced the mapping between the fingers and DSP stimuli and explained each of the identity cues prior to the pre-training session.

As each participant's training environment at home was different than the testing environment, arrangements were made to ease the transition to the testing environment (see \textbf{Figure \ref{fig1}} for the key layout during testing). Padding was placed under the participants' knees for comfort and participants were given a fiber optic response box with a configuration of buttons resembling that of the typical laptop used in training. For example, the distance between the centers of buttons in the top row was 20 mm (similar to the 20 mm between the "G" and "H" keys on a  MacBook Pro) and the distance between the top row and lower left button was 32 mm (similar to the 37 mm between the "G" and spacebar keys on a MacBook Pro). The position of the box itself was adjustable to accommodate participants' different reaches and hand sizes. In addition, padding was placed both under the right forearm to reduce strain during the task and also between the participant and head coil of the MRI scanner to minimize head motion.

Participants were tested on the same DSP task that they practiced at home, and, as in the training sessions, participants were given unlimited time to complete the trials with a focus on maintaining accuracy and responding quickly. Once a trial was completed, participants were notified with a ``+'' which remained on their screen until the next sequence-identity cue was presented. All sequences were presented with the same frequency to ensure a sufficient number of events for each type. Participants were given the same feedback after every ten trials as they were in training sessions. Each set of ten trials (referred to hereafter as trial blocks) belonged to a single exposure type (EXT, MOD, or MIN) and had five trials for each sequence, which were separated by an inter-trial interval that lasted between 0 and 6 seconds. Each epoch was composed of six blocks (60 trials) with 20 trials for each exposure and each test session contained five epochs and thus 300 trials.  Participants had a variable number of brain scans depending on how quickly they completed the tasks. However, the number of trials performed was the same for all participants, with the exception of two abbreviated sessions resulting from technical problems. In both cases, participants had only completed four out of five scan runs for that session when scanning was stopped. Data from these sessions are included in this study.

\textit{Behavioral apparatus}. The modules on participants' laptop computers were used to control stimulus presentation. These laptops were running Octave 3.2.4 along with the Psychtoolbox version 3. Test sessions were controlled using a laptop running MATLAB version 7.1 (Mathworks, Natick, MA). Key-press responses and response times were measured using a custom fiber optic button box and transducer connected via a serial port (button box, HHSC-1$\times$4-l; transducer, fORP932; Current Designs, Philadelphia, PA).

\begin{figure}
\centering
\includegraphics[width=100mm]{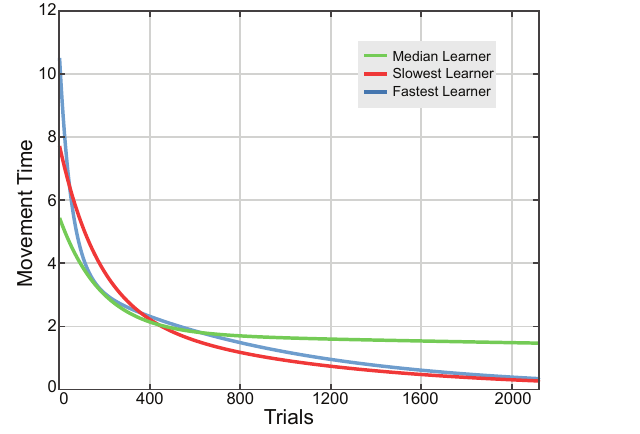}
\caption{\textbf{Exemplar Learning Curves.} To quantify learning rate, we fit a double exponential function to the movement time, defined as the time between the first button press and the last button press of a given sequence, as a function of trial number (see \textbf{Methods}). Fits are shown for the median learner, the fastest learner, and the slowest learner in the cohort. \label{learningComp}}
\end{figure}

\textit{Behavioral estimates of learning}. We defined movement time (MT) as the time between the first button press and the last button press for any single sequence. For the sequences of a single type, we fit a double exponential function to the MT \cite{schmidt1988motor, rosenbaum2009human} data in order to estimate learning rate. We used robust outlier correction in MATLAB (through the "fit.m" function in the Curve Fitting Toolbox with option "Robust" and type "Lar"): MT = $D_1e^{-t\kappa} + D_2e^{-t\lambda}$, where $t$ is time, $\kappa$ is the exponential drop-off parameter used to describe the fast rate of improvement (which we called the learning rate), $\lambda$ is the exponential drop-off parameter used to describe the slow, sustained rate of improvement, and $D_1$ and $D_2$ are real and positive constants. The magnitude of $\kappa$ determines the shape of the learning curve, where individuals with larger $\kappa$ values have a steeper drop-off in MT and thus are thought to be quicker learners (see \textbf{Figure \ref{learningComp}}) \cite{dayan2011neuroplasticity,snoddy1926learning}. This decrease in MT has been an accepted indicator of learning for several decades \cite{heathcote2000power} and various forms have been tried for the fit of MT \cite{newell1981mechanisms,heathcote2000power}, with variants of an exponential model being the most statistically robust choices. Importantly, this approach is also not dependent on an individual's initial performance or performance ceiling.

\subsection*{fMRI imaging.} \textit{Imaging procedures}. Signals were acquired using a 3.0 T Siemens Trio with a 12-channel phased-array head coil. Each whole-brain scan epoch was created using a single-shot echo planar imaging sequence that was sensitive to BOLD contrast to acquire 37 slices per repetition time (repetition time (TR) of 2000 ms, 3 mm thickness, 0.5 mm gap) with an echo time of 30 ms, a flip angle of 90 degrees, a field of view of 192 mm, and a 64$\times$64 acquisition matrix. Before the first round of data collection, we acquired a high-resolution T1-weighted sagittal sequence image of the whole brain (TR of 15.0 ms, echo time of 4.2 ms, flip angle of 90 degrees, 3D acquisition, field of view of 256 mm, slice thickness of 0.89 mm, and 256$\times$256 acquisition matrix).

\textit{fMRI data preprocessing}. Imaging data was processed and analyzed using Statistic Parametric Mapping (SPM8, Wellcome Trust Center for Neuroimaging and University College London, UK). We first realigned raw functional data, then coregistered it to the native T1 (normalized to the MNI-152 template with a resliced resolution of 3$\times$3$\times$3 mm), and then smoothed it with an isotropic Gaussian kernel of 8-mm full width at half-maximum. To control for fluctuations in signal intensity, we normalized the global intensity across all functional volumes. Using this pipeline of standard realignment, coregistration, normalization, and smoothing, we were able to correct for motion effects due to volume-to-volume fluctuations relative to the first volume in a scan run. The global signal was not regressed out of the voxel time series, given its controversial application to resting-state fMRI data \cite{murphy2009impact, saad2012trouble,chai2012anticorrelations} and the lack of evidence of its utility in analysis of task-based fMRI data. Furthermore, the functional connectivity matrices that we produce showed no evidence of strong global functional correlations but instead showed discrete organization in motor, visual and non-motor, non-visual areas \cite{bassett2015learning}. 

\subsection*{Network construction and analysis.} \textit{Partitioning into regions of interest} We divided the brain into regions based on a standardized atlas \cite{bassett2006small,braun2015human,bullmore2009complex}. There exist a number of atlases and the decision of which to use has been the topic of several recent studies on structural \cite{bassett2011conserved,zalesky2010whole}, resting-state \cite{wang2009parcellation}, and task-based network architectures \cite{power2011functional}. Consistent with prior graph-based studies of task-based fMRI \cite{bassett2011dynamic, bassett2013task,bassett2014cross, mantzaris2013}, we divided the brain into 112 cortical and subcortical regions using the Harvard-Oxford (HO) atlas of the FMRIB (Oxford Centre for Functional Magnetic Resonance Imaging of the Brain) Software Library \cite{macke2011biased, woolrich2009bayesian}. For each participant and for each of the 112 regions, the regional mean BOLD was computed by separately averaging across all voxels in that area (see \textbf{Figure \ref{fig8}} a \& b).

\textit{Wavelet decomposition}. Historically, wavelet decomposition has been applied to fMRI data \cite{bullmore2004wavelets,bullmore2003wavelets} to detect small signal changes in nonstationary time series with noisy backgrounds \cite{brammer1998multidimensional}. Here, we use the maximum-overlap discrete wavelet transform, which has been used extensively \cite{achard2006resilient, bassett2006adaptive, achard2007efficiency, achard2008fractal, bassett2009cognitive, lynall2010functional} to decompose regional time series into wavelet scales corresponding to specific frequency bands \cite{percival2006wavelet}. Because our sampling frequency was 2 s (1 TR), wavelet scale 1 corresponded to 0.125--0.25 Hz, and scale 2 to 0.06--0.125 Hz. To enhance sensitivity to task-related changes in BOLD magnitudes \cite{sun2004measuring}, we examined wavelet scale 2, consistent with our previous work \cite{braun2015dynamic, bassett2013task, bassett2014cross}. For a lengthier discussion of methodological considerations, see \cite{zhang2016choosing}.

\textit{Constructing Time-by-Time Networks}. We were interested in studying the similarities between brain states as individuals learn. We defined a brain state as a pattern of BOLD activity across brain regions at a single instant in time \cite{gu2015controllability, liu2013decomposition, liu2013time}. We measured the similarities between these states in each trial block which was comprised of approximately 40--60 repetition times (TRs).  We calculated the Spearman correlation of regional BOLD magnitudes between all possible pairs of time points (TRs). This procedure creates an undirected, weighted graph or network in which nodes represent time points and edges between nodes represent the correlation between brain states at different time points (see \textbf{Figure \ref{fig8}} c \& d). Intuitively, this matrix -- which we refer to as a ``time-by-time'' network provides the necessary information to uncover common brain states \cite{medaglia2015flexible,chen2016large}, and to study transitions between brain states, as a participant learns.

\textit{Isolating Brain States Using Community Detection}. To uncover common brain states in the ``time-by-time'' network, we used a network-based clustering technique known as \emph{community detection} \cite{fortunato2010,porter2009}. In particular, we chose a common community detection approach known as modularity maximization, where we optimize the following modularity quality function \cite{newman2006} using a Louvain-like \cite{blondel2008} locally greedy heuristic algorithm \cite{bassett2013robust}:
\[
Q_{0} = \sum_{ij} \left [ A_{ij}-\gamma P_{ij} \right ]\delta \left ( g_{i}, g_{j} \right )
\]

\noindent where $\mathbf{A}$ is the time-by-time matrix, times $\textit{i}$ and $\textit{j}$ are assigned respectively to community $g_{i}$ and $g_{j}$, the Kronecker delta $\delta \left ( g_{i}, g_{j} \right ) = 1$  if $g_{i}= g_{j}$  (and zero otherwise), $\gamma$  is the structural resolution parameter, and $P_{ij}$  is the expected weight of the edge between regions $\textit{i}$ and $\textit{j}$ under some null model. Consistent with prior work \cite{mattar2015,bassett2011dynamic,bassett2013task,bassett2015learning,braun2015dynamic,cole2014,mantzaris2013}, we used the Newman-Girvan null model \cite{girvan2002}:

\[
P_{ij} = \frac{k_{i}k_{j}}{2m}
\]

\noindent where $\textstyle k_{i} =\sum_{j} A_{ij}$ is the strength of region $\textit{i}$ and $\textstyle m = \frac{1}{2} \sum_{ij} A_{ij}$. Importantly, the algorithm we use is a heuristic that implements a non-deterministic optimization \cite{good2010}. Consequently we repeated the optimization 100 times \cite{bassett2013robust}, and we report results summarized over those iterations by building what are known as \emph{consensus partitions} \cite{bassett2015learning}  (see \textbf{Figure \ref{fig8}}). In order to do this, we construct a nodal association matrix $\mathbf{A}$ from a set of $N$ partitions, where $A_{i,j}$ is equal to the number of times in the $N$ partitions that node $i$ and node $j$ are in the same community. Furthermore, we construct a null nodal association matrix $\mathbf{A}^{n}$, constructed from random permutations of the $N$ partitions. This null association matrix indicates the number of times any two nodes will be assigned to the same community by chance. We then create the thresholded matrix $\mathbf{A}^T$ by setting any element $A_{i,j}$ that is less than the corresponding null element $A^n_{i,j}$ to $0$. This procedure removes random noise from the nodal association matrix $\mathbf{A}$. Subsequently, we use a Louvain-like method to obtain $N$ new partitions of $A^T$ into communities, where each of the $N$ partitions is typically identical, and each of which is a consensus partition of the $N$ original partitions.

\textit{Recurrent Brain States}.Each community obtained in the aforementioned pipeline includes a set of TRs that show similar patterns of regional BOLD magnitudes, and could thus be interpreted as representing a single, repeated brain state in a single trial bock. We first sought to aggregate these brain states over trial blocks. To this end, we average the pattern of regional BOLD magnitudes across all TRs assigned to that community in that trial block. We then repeat community detection across all representative brain states found in the trial blocks to find sets of representative brain states for each subject at each scan. By averaging the pattern of BOLD magnitudes of the brain states in each set, we find a group of representative brain states for every subject at every scan. \

Second, we sought to aggregate these subject-scan representative brain states over all scans to identify a group of representative brain states for each scan. We thus repeat community detection over the set of all subject-scan representative brain states, separated by scan and again average the pattern of regional BOLD magnitudes across all subject-scan representative brain states assigned to the same community. This final set of brain-states we consider to be scan-representative brain states for each scan of learning. \

Finally, we sought to find analogous communities in each scan. Therefore, we repeated the community detection algorithm for these communities and interpreted two scan-representative brain states assigned to the same community as analogous. In summary, we repeatedly use this brain state isolation procedure hierarchically to first isolate representative brain states for each subject-scan combination, then for each scan, and finally to find brain states in each scan that are similar to one another.

\section*{Results:}

\subsection*{Time by time network analysis identifies frontal and motor states} \

\begin{figure}
\centering
\includegraphics[width=130mm]{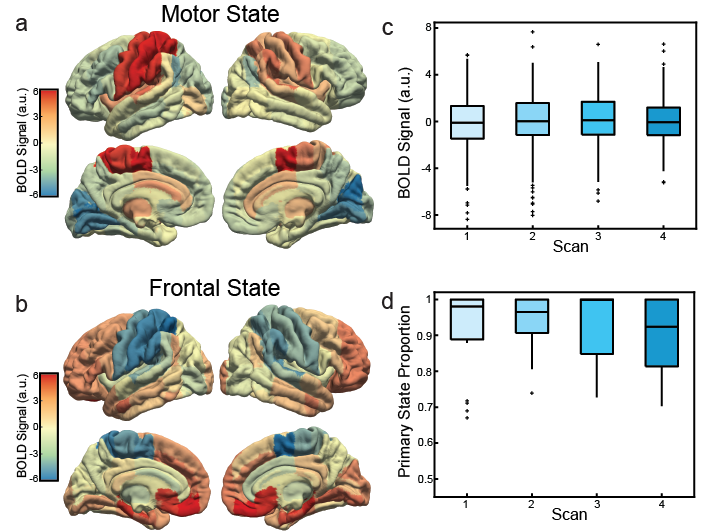}
\caption{\textbf{Brain States Common Across Learning.} We show the average activation magnitude of brain areas of the primary states. We refer to the first state shown in panel \emph{a} as the ``motor state'' due to the strong activation of the extended motor system. In this state, we also observe high activity magnitudes in anterior cingulate, and low activity magnitudes of primary visual cortex along the medial wall. We refer to the second state shown in panel \emph{b} as the ``frontal state'' due to the strong activation of frontal areas. In this state, we also observe high activation magnitudes in temporal cortex and subcortical structures, and low activity magnitudes in the extended sensorimotor system. In \emph{c} we show the average BOLD signal across all regions in each scan and note that it does not significantly change  ($F(3,669)=1.17, p=0.3221$). In \emph{d} we show that the primary state proportion does not significantly change across scans ($F(3,57)=0.17, p=0.9163$). \label{fig6}}
\end{figure}

Our first goal was to characterize the average anatomical distribution of BOLD magnitudes across all subjects and scans, to better understand the whole-brain activation patterns accompanying motor skill learning. To achieve this goal, we create a \textit{time-by-time network} where nodes represent individual time points, and edges represent the Spearman correlation coefficient between the vector of regional BOLD magnitudes at time point $i$ and time point $j$. We represented the time-by-time network as a graph. From these graphs, we were able to find 3 recurrent brain states, of which, two were strongly anti-correlated (Pearson correlation coefficient $r(446)=-0.4291$, $p=1.6951*10^{-21}$). These anti-correlated states make up 95.67\% of all time subjects spent learning, and are also the only states to be present in all scans. We therefore refer to these states as "primary states" and focus our analysis upon them. We refer to the first state as the ``motor state,'' characterized by strong activation of the extended motor system and anterior cingulate, as well as simultaneous deactivation of the medial primary visual cortex (Fig.~\ref{fig6} A, Table ~\ref{tab:states}). We refer to the second state as the ``frontal state,'' characterized by strong activation of a distributed set of regions in frontal and temporal cortices, as well as subcortical structures (Fig.~\ref{fig6} B, Table  \ref{tab:states}). 

\begin{table}[h!]
 \centering
 \caption {\textbf{Twenty regions with the greatest BOLD activity magnitude in the two primary brain states.}}
 \begin{tabular}{c c}  
 \hline
 Motor State & Frontal State\\ 
 \hline
 R, supplementary motor area &  L, frontal medial cortex \\ 
 L, supplementary motor area &  L, caudate \\
 L, postcentral gyrus &  R, frontal medial cortex \\
 L, superior parietal lobule &  L, parahippocampal gyrus, posterior \\
 R, planum temporale &  L, hippocampus \\ 
 L, supramarginal gyrus, anterior &  L, subcallosal cortex \\
 L, precentral gyrus &  R, caudate \\
 R, supramarginal gyrus, anterior &  R, parahippocampal gyrus, posterior \\
 R, precentral gyrus &  R, subcalosial cortex \\
 R, superior parietal lobule &  R, hippocampus \\
 R, parietal operculum cortex &  R, parahippocampal gyrus, anterior \\
 L, Heschl's gyrus &  R, nucleus accumbens \\
 L, parietal operculum cortex &  L, middle temporal gyrus, anterior \\
 L, globus pallidus &  L, parahippocampal gyrus, anterior \\
 R, central opercular cortex &  L, paracingulate gyrus \\
 R, supramarginal gyrus, posterior &  R, lingual gyrus \\
 R, Heschl's gyrus &  L, inferior frontal gyrus, pars triangularis \\
 L, central opercular cortex &  L, parahippocampal gyrus (superior to ROIs 34,35) \\
 L, planum temporale &  L, nucleus Accumbens \\
 R, postcentral gyrus &  R, middle temporal gyrus, anterior \\
 \hline
 \end{tabular}  \label{tab:states}
\end{table}

While these two states were statistically present across the entire experiment, we did observe small fluctuations in the magnitudes of the regional activity of both states. Thus, natural questions to ask are (i) did either state became stronger or weaker with training? and (ii) did the frequency of primary states change with learning? To address the first question, we calculated the mean BOLD magnitude among all brain regions for each state. Using a repeated measures ANOVA, we found no significant differences among scans in either state ($F(3,669)=1.17, p=0.3221$ (\textbf{Figure \ref{fig8}}). This suggests that the activation of these two states did not significantly change -- on average -- with the level of training. To address the second question, we calculated the proportion of all states that the primary states make up in each scan. Using a repeated measures ANOVA, we found no significant differences among scans ($F(3,57)=0.17, p=0.9163$). This suggests that the frequency of primary states remained the same during learning.

\subsection*{State flexibility increases with task practice} \

\begin{figure}
\centering
\includegraphics[width=80mm]{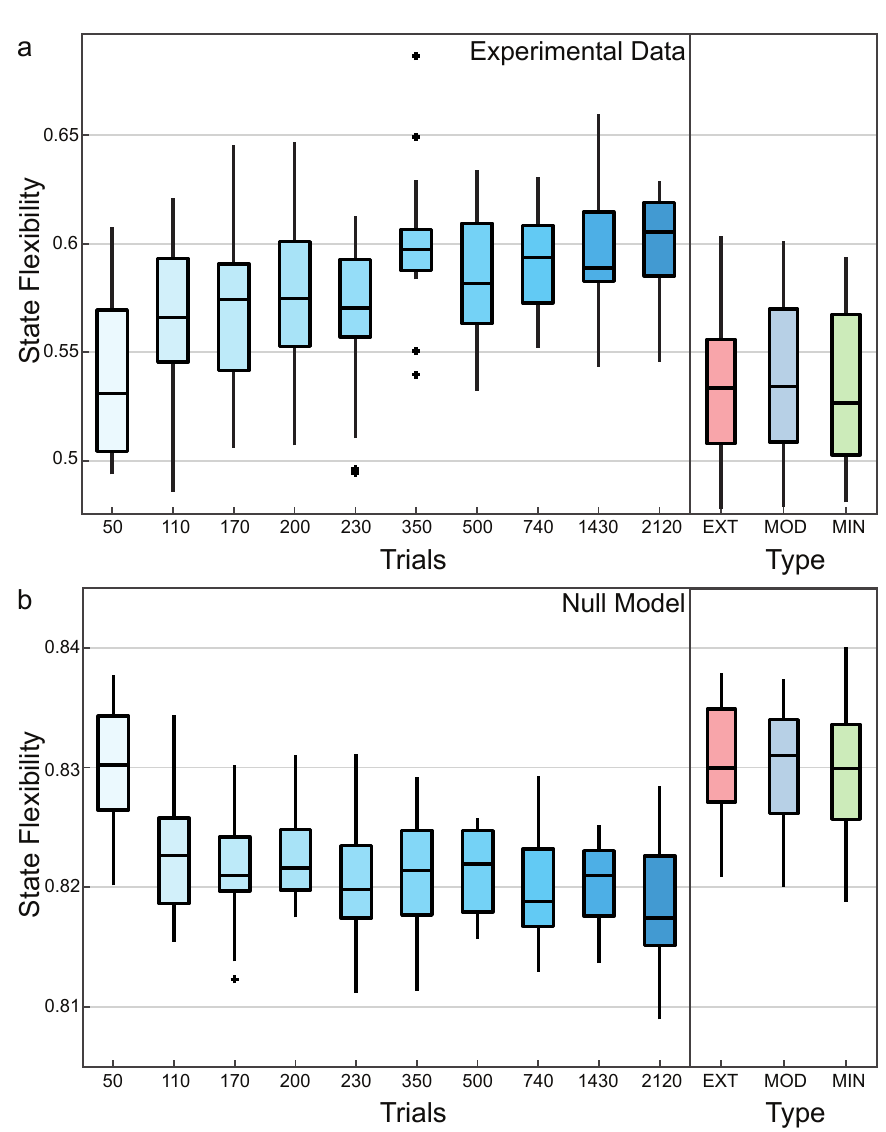}
\caption{\textbf{Whole-Brain State Flexibility.} \emph{(a) Left Panel:} State flexibility shows a clear increasing trend as subjects complete more trials of the experiment ($F(9, 171)=9.97, p=3.0417 \times 10^{-12}$ using a repeated measures ANOVA). \emph{Right Panel:} State flexibility in pre-training scans for sequences that would become extensively trained (EXT), sequences that would become moderately trained (MOD) and sequences that would become minimally trained (MIN). There was no effect of sequence type on state flexibility. These three sets of data compose the 50 trial boxplot in the left panel. \emph{(b) Left Panel:} State flexibility shows a mild decreasing trend in the null model due to the reduced number of TRs ($F(9, 171)=2.6, p=0.0078$; null model was created by permuting the adjacency matrix while maintaining symmetry, see Supplementary Materials for more detail). \emph{Right Panel:} State flexibility in pre-training scans for sequences that would become extensively trained (EXT), sequences that would become moderately trained (MOD) and sequences that would become minimally trained (MIN). These three sets of data compose the 50 trial boxplot in the left panel. These results indicate that we find no effect of sequence type prior to training. \label{fig2}}
\end{figure}

How does the brain traverse these states? Do individuals' traversals change with learning? To examine how the pattern of traversals through brain states changes during learning, we defined a ``state flexibility'' metric. Following \cite{medaglia2015flexible}, we specified state flexibility $(F)$ to be the number of state transitions $(T)$ observed relative to the number of states $(S)$, or $F= \frac{T}{S}$. Intuitively, state flexibility is a measure of the volatility versus rigidity in brain dynamics, directly representing the frequency of dynamic state changes. We observed that state flexibility increased monotonically with the number of trials practiced (repeated measures ANOVA: $F(9, 171)=9.97, p=3.0417 \times 10^{-12}$, \textbf{Figure \ref{fig2}}). This suggests that as subjects learned the sequences, regional patterns of BOLD magnitudes became more variable, indicating more frequent transitions between different brain states.

An important question to consider is whether this change in state flexibility is related to the length of time that participants take to complete the practice trials. Specifically, because the experiment is self-paced, the length of time to complete the sequences decreased as participants practiced; subjects became quicker with experience. To ensure that the length of time to complete a sequence was not driving the observed changes in state flexibility, we constructed a non-parametric permutation-based null model by permuting the adjacency matrix $\mathbf{A}$ uniformly at random while maintaining symmetry. Critically, this null model displayed a \emph{decrease} in state flexibility with number of trials practiced ($F(9, 171)=2.6, p=0.0078$, \textbf{Figure \ref{fig2}}), suggesting that neither the reduced length of time nor the correlation values themselves can explain the observed increase in state flexibility, but rather that the temporal structure of the data is required for the observed increase in state flexibility.

\subsection*{Regional contributions to state flexibility vary by function} \

Do regions contribute differentially to state flexibility? To answer this question we conducted a ``lesioning'' analysis, where we calculated state flexibility for each subject and scan while ``lesioning out,'' or excluding, a single region. We then calculated the average difference between the true state flexibility for that subject and scan, and the lesioned state flexibility for each region across all trial groups. We normalized these values by subtracting off the mean effect of lesioning on flexibility.

\begin{figure}
\centering
\includegraphics[width=140mm]{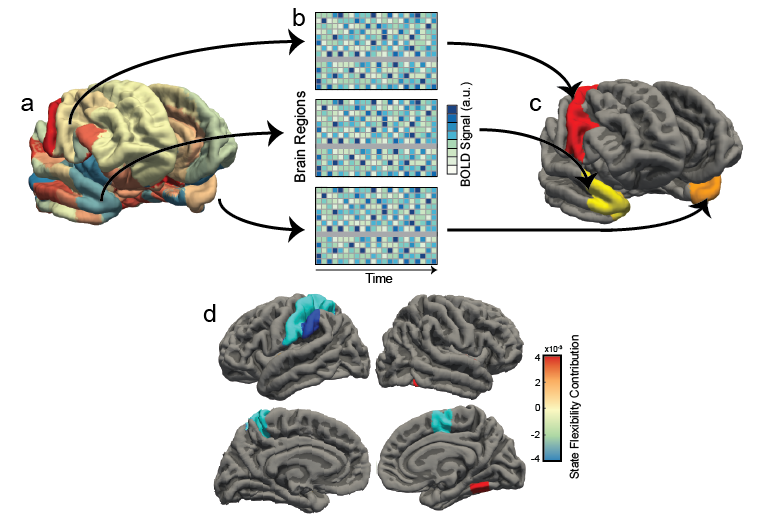}
\caption{\textbf{Schematic Depicting Ablasion Calculation and Regional Contribution to State Flexibility.} \emph{(a)} Blood-oxygen-level-dependent (BOLD) signal from functional magnetic resonance imaging (fMRI) data was acquired from healthy adult subjects. \emph{(b)} Single regions were selected and ``ablated'' or ignored in the BOLD time-series. \emph{(c)} State flexibility was recalculated for these new matrices and the difference between true state flexibility and the ablated state flexibility is defined to be the regional contribution to state flexibility for the ablated region. \emph{(d)} We show the difference between the true state flexibility and ablated state flexibility for each region. We note that regions in the motor and visual cortex tend to have negative contributions to state flexibility while regions in the frontal lobe tend to have positive contributions to state flexibility.\label{fig7}}
\end{figure}

To assess statistical significance, we created a matrix of the contributions to state flexibility for all regions and subjects. We found the contribution from seven regions to be significant ($p<0.05$ for each region, $df = 19$) by computing a $t$-test between the ablated state flexibilities and the true state flexibility, while correcting for multiple comparisons across the 112 brain regions using the false discovery rate. By calculating the average of these contributions for each region, we identify negative contributors, the removal of which increases state flexibility, and positive contributors, the removal of which decreases state flexibility. We find that the significant negative contributors to state flexibility are associated generally with motor and visual function (supplementary motor area, cuneus cortex, and the postcentral gyrus). In contrast, the significant positive contributors to state flexibility are associated with more integrative processing in hetermodal association areas (temporal occipital fusiform cortex, and planum polare on the temporoparietal junction) (Fig.~\ref{fig7}, Table~\ref{tab:abl}).

\begin{table}[h!]
 \centering
 \caption {\textbf{Significant contributors to state flexibility.}}
 \begin{tabular}{c c}  
 \hline
 Most Negative Contributors & Most Positive Contributors\\ 
 \hline
R, supplementary motor area &	L, Planum polare \\
L, Superior parietal lobule &	R, Planum polare \\
L, Postcentral gyrus	& R, Temporal occipital fusiform cortex \\
L, Supramarginal gyrus, anterior	\\
 \hline
 \end{tabular}  \label{tab:abl}
\end{table}

\subsection*{State flexibility is correlated with learning rate} \

The results thus far indicate that state flexibility is an important global feature of brain dynamics that significantly changes as individuals learn a new motor-visual skill. Yet, they do not address the question of how such brain dynamics relate directly to changes in behavior. Therefore, we next asked the question: are individual differences in state flexibility related to individual differences in learning rate (as defined in Methods)? Here, we focus solely on the most trained sequences, as the effects of learning are most dramatic in the most frequently practiced sequences \cite{bassett2015learning, wymbs2015human}. Here, we estimate the correlation between the learning rate between sessions, and the differences in state flexibility between sessions. All correlations are estimated using a mixed linear effects model that accounts for the effect of either subject or scan.

\begin{figure*}
\centering
\includegraphics[width=80mm]{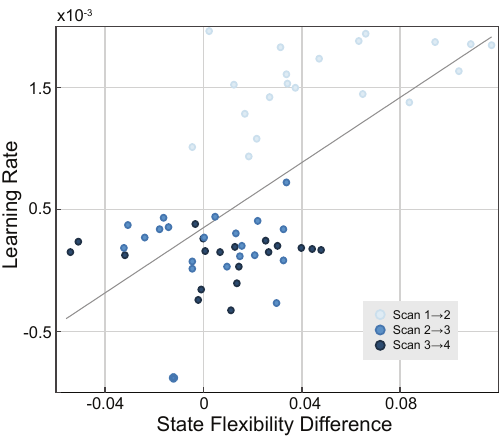}
\caption{\textbf{Individual Differences in Learning Rate Correlated with State Flexibility.} State flexibility difference refers to the difference in state flexibility between subsequent scans. State flexibility differences for all regions were computed and were found to be significantly positively correlated with individual differences in learning rate ($p=1.163\times10^{-7}$), suggesting that the observed increase in flexibility is associated with the learning rate of subjects.\label{fig4}}
\end{figure*}

Using this method, we observe a significant positive correlation between state flexibility and learning rate ($p=1.163\times10^{-7})$, (\textbf{Figure \ref{fig4}}), accounting for the effects of subject.That is, subjects will tend to be inherently better or worse at learning than other subjects, therefore, we normalize for inter-subject differences in learning rate, and we find a significant correlation between  state flexibility difference and learning rate. State flexibility difference is positively correlated with learning, and thus larger decreases in flexibility are associated with better learning. Furthermore, we observe a significant correlation between state flexibility difference and learning rate ($p=0.0376$) when accounting for the effects of scan. Learning rate tends to decrease as the number of scans increases, therefore, we build this trend into our model, and find a significant correlation between  state flexibility and learning rate. Extending our previous assertion, these results suggest that both the individual differences and the larger patterns of change are correlated with learning rate.

\section*{Discussion}

In this work, we studied task-based fMRI data collected at 4 time points separated by about 2 weeks during which healthy adult participants learned a set of six 10-note finger sequences. During learning, we hypothesized that the brain would show a change in the manner in which it traversed brain states. We defined a state as a pattern of BOLD magnitude across 112 anatomically-defined brain regions. We identified two canonical states characteristic of the entire period of task performance, which showed high activation of motor cortex and frontal cortex, respectively. Interestingly, we observed that the flexibility with which participants switched among these canonical states and other less common states was lowest early in training and highest late in training, indicating the emergence of state flexibility. We find that the positive contributors to state flexibility are associated with integrative processing while the negative contributors are associated with motor and visual function. Finally, we observe that changes in state flexibility were correlated with learning rate: increasing state flexibility was correlated with higher learning rates. 

\subsection*{Extensions of graph theoretical tools to the temporal domain}

Over the past decade, tools from graph theory have offered important insights into the structure and function of the human and animal brain, both at rest and during cognitively demanding tasks \cite{sporns2010}. In these applications, the nodes of the graph are traditionally thought of as neurons or brain areas, and the edges of the graph are defined by either anatomical tracts \cite{hagmann2008mapping,bassett2011conserved} or by functional connections \cite{achard2006resilient,bassett2006adaptive}. Yet, the tools of graph theory are in fact much more general than these initial applications \cite{bollobas1979graph,bollobas1985random}. Indeed, recent extensions have brought these tools to other domains -- from genetics \cite{fulcher2016transcriptional,conaco2012functionalization,arcila2014novel} to orthopedics \cite{murphy2016structure} -- by carefully defining alternative graph representations of relational data. As a concrete example, a graph can be used to encode the relationships between movements or behaviors, by treating a movement as a node, and by linking subsequent movements (or actions) by the inter-movement interval \cite{wymbs2012differential}. Similarly, a graph can be used to code the temporal dependencies between stimuli, by treating a stimulus as a node, and by linking pairs of stimuli by their temporal transition probabilities \cite{schapiro2013neural,karuza2016process}.

While these applications may initially seem vastly different, they in fact all share a common property: that entities are related to one another by some facet of time. Here, by contrast, we construct the edge-vertex dual of this more common form. We ask: How are times related to one another by some other entity? Specifically, we study how the brain state in one time point is related to the brain state in another time point, and we define a brain state as the vector of activation magnitudes across all regions of interest \cite{medaglia2015flexible,chen2016large}. The notion that a pattern of activation reflects a brain state is certainly not a new one \cite{gu2015controllability}. In the context of fMRI data, a common approach is to study the multi-voxel pattern of activation in a region of interest to better understand the representation of a stimulus \cite{kubilius2015brain,chadwick2012decoding}. And in the context of EEG and MEG data, the pattern of power or amplitude in a set of sensors or a set of reconstructed sources is frequently referred to as a microstate \cite{wegner2016analytical}. The composition and dynamics of these microstates have shown interesting cognitive and clinical utility, predicting working memory \cite{muthukrishnan2016functional} and disease \cite{gschmwind2016fluctuations}. Yet, while patterns of activation are acknowledged as an important representation of a brain or cognitive state, little is known about how these states evolve into one another. Recent advances have made this possible by coding the relationships between brain states in a graph \cite{medaglia2015flexible}. Here we capitalize on these advances to extract the community structure in such a graph, to identify canonical states, and to quantify the transitions between them. It will be interesting in future to broaden the analytical framework applied here to study other properties of the graph -- including local clustering and global efficiency -- to better understand how the brain traverses states over time.

\subsection*{Brain states characteristic of discrete sequence production}

Using this unusual graph theory approach in which network nodes represent time points and network edges represent similarities in brain states across two time points, we were able to identify two canonical brain states that characterized the task-evoked activity dynamics across the entire experiment, extending across 6 weeks of intensive training. The most common state, perhaps unsurprisingly, was characterized by high BOLD magnitudes in regions of the extended motor cortex, including the bilateral precentral gyrus, left postcentral gyrus, bilateral superior parietal lobule, bilateral supramarginal gyrus, bilateral supplementary motor area, bilateral parietal operculum cortex, and bilateral Heschl's gyrus \cite{dayan2011neuroplasticity}. This map is consistent with the fact that this is an intensive motor-learning paradigm \cite{bassett2015learning,wymbs2015human} in which participants acquire the skill necessary to perform a sequence of 12 finger movements over a short period of time. The second most common state was composed of a frontal-temporal-subcortical system, containing the anterior middle temporal gyrus, medial frontal cortex, parahippocampal gyrus, caudate, nucleus accumbens, and hippocamus. These areas are thought to play critical roles in sequence learning \cite{exner2002differential,vakil2000motor} facilitated by higher-order cognitive processes including reward learning \cite{hikosaka2014basal,kim2015parallel}, cognitive control and executive function \cite{stuss2011functions,alvarez2006executive}, predicting nature and timing of action outcomes \cite{brown2011medial,grinband2011dorsal,rushworth2005cognitive}, and subcortical storage of motor sequence information \cite{lehericy2005distinct}. This system is particularly interesting because it displayed a competitive relationship with the motor state, with a strongly anticorrelated activation profile, suggesting that frontal-subcortical circuitry affects control by transient, desynchronized interactions.

\subsection*{State flexibility, task practice, and learning rate}

Beyond the anatomy of the states that characterize extended training on a discrete sequence production task, it is also useful to study the degree to which those states are expressed, and the manner in which one state moves into another state. The two primary states that we observed characterized $95.67\%$ of all time points, indicating their canonical nature. Temporally, the brain frequently switched back and forth between these two states, with less frequent traversal of other non-primary states. We quantified this switching using a brain state flexibility measure \cite{medaglia2015flexible}, and observed that flexibility increased significantly over the course of the 6 weeks of training. Moreover, brain state flexibility was negatively correlated with learning rate, being lowest early in training when behavioral adaptivity was greatest. These results suggest that consistent activation patterns characterize early training, when participants must learn the mapping of visual cues to motor responses, the use of the button box, and the patterns of finger movements. Later in learning, when the skill has become relatively automatic, participants display more varied progressions of activation patterns (higher brain state flexibility), potentially mirroring the greater freedom of their cognitive resources for other processes \cite{shamloo2016changes}. 

Importantly, these results offer a complement to prior efforts to quantify network flexibility based on estimates of functional connectivity \cite{bassett2011dynamic} where the nodes are brain regions and the edges are temporally defined correlations between those regions. Network flexibility appears to peak early in finger sequence training \cite{bassett2011dynamic}, followed by a growing autonomy of motor and visual systems \cite{bassett2015learning}. In combination with our results, these prior data suggest that there may be distinct time scales associated with brain variability at the level of activity (where variability may peak late) in comparison to the level of connectivity (where variability may peak early). Such a hypothesis could be directly validated in additional studies that reproduce the results we present here. The apparent separation in time scales of these processes over learning also supports the growing notion that the information housed in patterns of activity can be quite independent from information housed in patterns of connectivity \cite{bassett2015learning,cao2016functional,seibenhuhner2013}. For example, earlier studies have demonstrated that patterns of beta weights from a GLM do not necessarily map onto patterns of strong or weak functional connectivity \cite{bassett2015learning}, the temporal dynamics of an activity time trace do not necessarily map onto patterns of functional connectivity \cite{seibenhuhner2013}, and phenotypes indicative of psychiatric disease can be identified in functional connectivity while being invisible to methods focused on activity \cite{esslinger2009neural}. Together, these studies indicate that activity and connectivity can provide \emph{distinct} information regarding the neurophysiological processes relevant for cognition and disease. They also in principle support the possibility of differential time scales of flexibility in activity and connectivity as a function of learning. 

\subsection*{Methodological Considerations}

There are several important methodological and conceptual considerations pertinent to this work. The first consideration we would like to discuss is a relatively philosophical one. It pertains to our use of the term ``brain state''. It is important to disambiguate the use of \emph{brain state} as a quantifiable and quantified object, defined as the pattern of activation magnitudes over all brain areas (strung out in a vector \cite{gu2015controllability}), and other more conceptual notions of mental state or cognitive state. These latter notions can be difficult to quantify directly from imaging data, even if they may have relatively specific definitions from both psychological and clinical perspectives \cite{richman2015mental,martin2016what,eddy2016changes}. It will be important in future uses of our brain state detection and characterization technique to maintain clarity in the use of these terms.

The second important consideration relevant to this work, is that the data that we study here was collected with a tranditional 2 second TR. It would be very interesting to test for similar phenomena in the high-resolution BOLD imaging techniques available now, for example using multiband acquisitions. Such higher sampling could offer heightened sensitivity to changes in brain state flexibility related to individual differences in learning. Moreover, they could provide enhanced sensitivity to variations in brain state flexibility across different frequency bands, particularly higher frequency bands that have been shown to be sensitive to shared genetic variance \cite{fornito2011genetic}.

Finally, on a computational note, it is important to emphasize that the results described here are obtained via the application of a clustering technique \cite{porter2009,fortunato2010} to identify brain states from the temporal graph. Importantly, the technique that we use -- based on modularity maximization \cite{newman2006} -- is a hard partitioning algorithm that seeks to solve an NP-hard problem using a clever heuristic \cite{blondel2008}. Although modularity maximization can accurately recover planted network modules in synthetic tests \cite{lancichinetti2009community,bassett2013robust}, it does have important limitations \cite{lancichinetti2011,good2010}. Therefore it would be interesting in future to examine the sensitivity of results to other clustering techniques available in the literature.

\section*{Conclusion}

In summary, in this study we seek to better understand the changes in brain state that accompany the acquisition of a new motor skill over the course of extended practice. We treat the brain as a dynamical system whose states are characterized by a recognizable pattern of activation across anatomicaly defined cortical and subcortical regions. We apply tools from graph theory to study the temporal transitions (network edges) between brain states (network nodes). Our data suggest that the emergence of automaticity is accompanied by an increase in brain state flexibility, or the frequency with which the brain switches between activity states. Broadly, our work offers a unique perspective on brain variability, noise, and dynamics \cite{deco2009key,breakspear2011networks,garrett2013moment,garrett2014brain}, and its role in human learning.

\section*{Acknowledgments} DSB would like to acknowledge support from the John D. and
Catherine T. MacArthur Foundation, the Alfred P. Sloan Foundation, the Army Research Office through contract number W911NF-14-1-0679, the National Institute of Health (1R01HD086888-01), and the National Science Foundation (BCS-1441502, CAREER PHY-1554488, BCS-1631550, and CNS-1626008). The content is solely the responsibility of the authors and does not necessarily represent the official views of any of the funding agencies.

\bibliography{bibfile_v1}
\bibliographystyle{abbrv}

\newpage
\end{document}